\documentclass[aip, amsmath,amssymb,
 reprint,%
]{revtex4-1}

\usepackage{dcolumn}
\usepackage{bm}
\usepackage{amsmath}
\usepackage{braket}
\usepackage{graphicx}
\usepackage{float}

\usepackage[utf8]{inputenc}
\usepackage[T1]{fontenc}
\usepackage{mathptmx}
\usepackage{etoolbox}
\usepackage{ulem}

\usepackage{xcolor}

\newcommand{\um}{\ensuremath{\mu}m}
\newcommand{\us}{\rm\ensuremath{\mu}s}
\newcommand{\none}{{ --}}
\newcommand{\jonathan}[1]{{\color{black}#1}}
\newcommand{\kater}[1]{{\color{black}#1}}

\makeatletter
\def\@email#1#2{%
 \endgroup
 \patchcmd{\titleblock@produce}
  {\frontmatter@RRAPformat}
  {\frontmatter@RRAPformat{\produce@RRAP{*#1\href{mailto:#2}{#2}}}\frontmatter@RRAPformat}
  {}{}
}%
\makeatother
\begin{document}
	\title{Optical Direct Write of Dolan\kater{--Niemeyer}-Bridge Junctions for Transmon Qubits}

	\author{J. T. Monroe}

	\author{D. Kowsari}

	\author{K. Zheng}
	
	\author{C. Gaikwad}
	
	\affiliation{Department of Physics, Washington University, St. Louis, Missouri 63130, USA}
	
	\author{J. Brewster}	\affiliation{Department of Physics, Saint Louis University, St. Louis, Missouri 63103, USA}
	
	\author{D. S. Wisbey}
	\affiliation{Department of Physics, Saint Louis University, St. Louis, Missouri 63103, USA}
	
	\author{K. W. Murch}
	\email[]{Authors to which correspondence should be addressed: murch@physics.wustl.edu, j.monroe@wustl.edu }
	\affiliation{Department of Physics, Washington University, St. Louis, Missouri 63130, USA}

	\date{\today}
	
	\begin{abstract}
		We characterize highly coherent transmon qubits fabricated with a direct-write photolithography system. Multi-layer evaporation and oxidation allows us to \kater{change the critical current density by}  reducing the effective tunneling area and increasing the barrier thickness. \kater{Surface treatments before resist application and again before evaporation result in high coherence devices.}    With optimized surface treatments we achieve energy relaxation $T_1$ times in excess of $80\ \mu$s for three dimensional transmon qubits with Josephson junction lithographic areas of 2 $\mathrm{\um^2}$.
	\end{abstract}
	
    \pacs{}
	
	\maketitle

		Transmon qubits are a leading platform employed in noisy intermediate scale quantum processors \cite{Otterbach2017, Arute2019, Harrigan2021, Kjaergaard2020}.
	The success of these devices has resulted from continued advances in fabrication techniques, with state of the art cm-scale devices incorporating mm-, $\mu$m-, and nm-scale features. Typical fabrication utilizes multiple lithography and metalization steps. While Josephson junctions are fabricated at the smallest length scales, particularly to reduce the presence of two-level-system (TLS) fluctuators \cite{Martinis2005,Stoutimore2012}, other circuit elements are fabricated with larger feature sizes to reduce coupling \cite{Barends2010,Geerlings2012,Jin2017,Niepce2020} to TLS fluctuators  \cite{Muller2017,McRae2020}.  We now understand that TLS defects primarily reside at substrate-metal, metal-metal, and metal-air interfaces \cite{Gao2008,Wisbey2010,Lisenfeld2019,Bilmes2020,Grunhaupt2017,Dunsworth2017,Osman2020}.  Careful surface cleaning has vastly improved coherence times \cite{Zeng2015,Nersisyan2019a, Altoe2020}.

	Fabrication based exclusively on photolithography simplifies processing
	and reduces the need for multiple lithography steps \cite{Foroozani2019} \kater{decreasing fabrication time and cost and enabling larger scale production}.
	Photolithography has shown consistency and uniformity across wafers, yet is limited by its relatively large minimum feature size.
	Although advanced lithography techniques offer workarounds to improve the resolution limit \cite{Lin1987,Wagner2000,Bencher2008,Miyaji2016}, 
	standard photolithography cannot \kater{reliably} pattern features as small as typical Josephson junctions (JJs) \kater{for qubits, which have features on the scale of 100 nm}.
	
	
	In this Letter, we demonstrate that highly coherent transmon qubits can be fabricated with an all-optical direct-write photolithography system.
	Direct writing allows fast prototyping by rastering a focused optical beam across a resist-coated substrate.
	\jonathan{We test two oxidation procedures that both produce appropriate critical currents, despite large overlap areas. First, a multi-step oxidation process creates small effective area JJs. Second, a single-step oxidation creates JJs with large effective areas. Transmon qubits with both oxidation procedures exhibit high $T_1$ times, but our highest observed $T_1$ times come from single-step-oxidation devices. }	Moreover, we demonstrate the need for multiple cleaning steps to achieve low loss.
	
	This work joins a growing body of literature studying alternative fabrication techniques for transmon qubits.
	Other work using photolithography \cite{Foroozani2019,Kim2021}  demonstrated modest coherence times, and merged-element transmons have used large overlap areas in compact footprints \cite{Zhao2020,Mamin2021}.
	Our work extends elements of these studies, using photolithography to make large overlap area JJs for highly coherent three dimensional transmon qubits.

	We fabricate large-area JJs with aluminum double-angle-evaporated onto high-resistivity, 100-oriented silicon substrates. 	
	Our best devices result from substrate cleaning for 10 minutes in a 3:1 mixture of sulfuric acid 
	and hydrogen peroxide heated to $120^\circ$ C (Piranha etch) 
	 followed by a 5-minute buffered-oxide etch (BOE). 
These cleaning steps remove native silicon oxides and residual organics from the silicon surface \cite{Zeng2015a,Nersisyan2019a, Altoe2020,Place2021}. 

\begin{figure}
		\centering
		\includegraphics[width=0.9\linewidth]{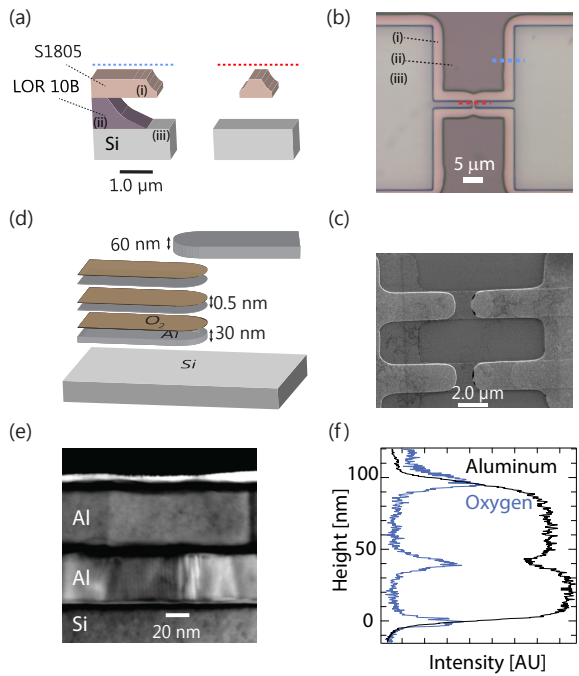}
		\caption{Fabrication method for large-area JJs. 
			(a) A sketch of the bi-layer resist stack. The 1.5 ${\rm \mu m}$ undercut creates a Dolan bridge to mask double-angle-deposited aluminum. Roman numerals and dashed lines indicate resist coverage depicted in panel b.
			(b) An optical image of the post-development resist stack indicates undercut regions. 
			(c) A scanning electron microscope image of two large-area JJs in a SQUID geometry. 
			(d) A multi-layer oxidation process achieves a thick oxide for low critical current JJs. 
			(e) Cross-sectional TEM image of the oxidation stack. 
			(f) Averaged EDXS  abundance of aluminum and oxygen. Oxygen in the middle of the aluminum region defines the  tunneling barrier. }
		\label{figStack}
	\end{figure}

	We design a bi-layer optical resist stack to support Dolan-Niemeyer-bridge shadow mask evaporation \cite{Dolan1977,Niemeyer1976}. 
	The bottom layer is a 
	liftoff photoresist (MicroChem LOR 10B). The top layer is a high-resolution imaging photoresist (MicroChem Shipley S1805). We spin and softbake these resists to achieve 1 \um~height in the liftoff resist and 0.6 \um~in the imaging resist, as illustrated in Figure \ref{figStack}(a).
	Our recipes are tailored for 1.5 \um~undercuts, but the liftoff photoresist supports undercuts up to 10 \um. 
	
	An all-optical 375 nm direct-write-lithography system (Heidelberg DW 66+) exposes the resists by rastering a focused optical beam across the wafer. 
	Figure \ref{figStack}(b) shows a patterned 1.5 \um~long and 0.8 \um~wide Dolan bridge. The dimensions are chosen to stay above the lithography system's resolution limit.
	Scanning electron microscope (SEM) imaging shows a typical JJ with an area of $1.5~\mu \mathrm{m}^2$ (Fig.~\ref{figStack}c). These overlap areas are two orders of magnitude larger than JJs fabricated with electron-beam lithography that have areas $\sim 0.01~\mu \mathrm{m}^2$. 
	We develop exposed resist in a metal-ion-free developer (MicroChem MF-319) and visually confirm the Dolan bridge geometry (Fig.~\ref{figStack}b). 
	
	After development, we employ oxygen plasma ashing (Plasma Etch PE 50, 20s, 100 W, O$_2$ 15 cc/min) to remove any residual resist \cite{Kreikebaum2019}. A 30-second BOE step removes silicon oxides that have grown on the atmosphere-exposed substrate \cite{Altoe2020}.  
	The wafer is rapidly transferred ($<$5 minutes) to an ultra-high vacuum ($\sim$5 nTorr) environment. The sample remains at low pressure for $\sim$18 hours, in order to further remove any adsorbed contaminants.
	We then evaporate five-nine purity aluminum onto the substrate using the Dolan bridge shadow-mask technique \cite{Dolan1977}. This evaporation consists of three primary steps: i) a 30 nm layer of aluminum is evaporated at $45^\circ$ relative to the wafer's normal vector at a rate between 0.3-1 nm/s; ii) an oxidation step, detailed below, creates \kater{the AlO$_x$}  layer for the JJ; iii) a 60 nm layer of aluminum is evaporated at $-45^\circ$, forming the JJ.
	
	
	Achieving low critical currents with large-overlap-area JJs requires increasing the oxide thickness.  \kater{Here we discuss a multi-step oxidation procedure to grow a sufficiently thick oxide layer, sketched in Fig.~\ref{figStack}(d), and below we also present results based on a single oxidation step.}
	The recipe leverages the fast initial growth of the thin-film oxides predicted by Cabrera-Mott theory \cite{Cabrera1949}.
	After evaporating the initial 30 nm of aluminum, we grow an oxide film with 99.9\% pure oxygen at 4.3 Torr for 300 seconds. We then evaporate a variable number of 0.5 nm aluminum filler layers at the same evaporation angle, each oxidized under the same conditions. Multiple filler layers are usually needed to achieve suitable JJ resistances.
	
	Cross-sectional transmission electron microscope (TEM) images (Fig.~\ref{figStack}(e)) and energy dispersive X-ray spectroscopy (EDXS)   indicates a 4.5 nm oxide-barrier thicknesses for the multi-step oxidation (Fig.~\ref{figStack}(f)).
	We note that this estimate is only a rough upper bound because of averaging over significant surface roughness.
	We also note that TEM imaging probes structural rather than electrochemical properties, thus overestimating the effective barrier thickness \cite{Kim2020a}.
	Nonetheless, this oxide barrier is significantly thicker than typical single-layer JJs \cite{Zeng2015, Olsson2016}. After liftoff of sacrificial aluminum in N-Methyl-2-pyrrolidone (NMP), we probe samples with a DC voltage to infer the critical current via the Ambegaokar-Baratoff relation \cite{Ambegaokar1963}. 

	We embed two large-area JJs into a SQUID geometry with a shunt capacitor of $\sim$60 fF, typical of three-dimensional transmons \cite{Paik2011}. We place the circuit into a copper cavity and cool it in a dilution refrigerator to 10 mK. A solenoid mounted to the cavity exterior generates a DC magnetic field through the SQUID loop to tune the circuit's resonant frequency.
	
	Two-tone spectroscopy is used to characterize the energy structure of the circuit. At high probe power, the probe excites multi-photon transitions.
	We measure the circuit's anharmonicity and infer the Josephson energy, $E_J$, and charging energy, $E_C$, from the transmon Hamiltonian. For the devices in this study, the ratio $E_J/E_C$ is between 20 and 70, well within the transmon regime \cite{Koch2007}.
	Despite the $\sim$100-fold increase in junction area, the charging energy \kater{of the circuit} remains comparable to \kater{a circuit containing only} small-area junctions.	We attribute this to the presence of two {JJ}s, detailed below.
	
	\begin{figure}[ht]
		\centering
		\includegraphics[width=0.9\linewidth]{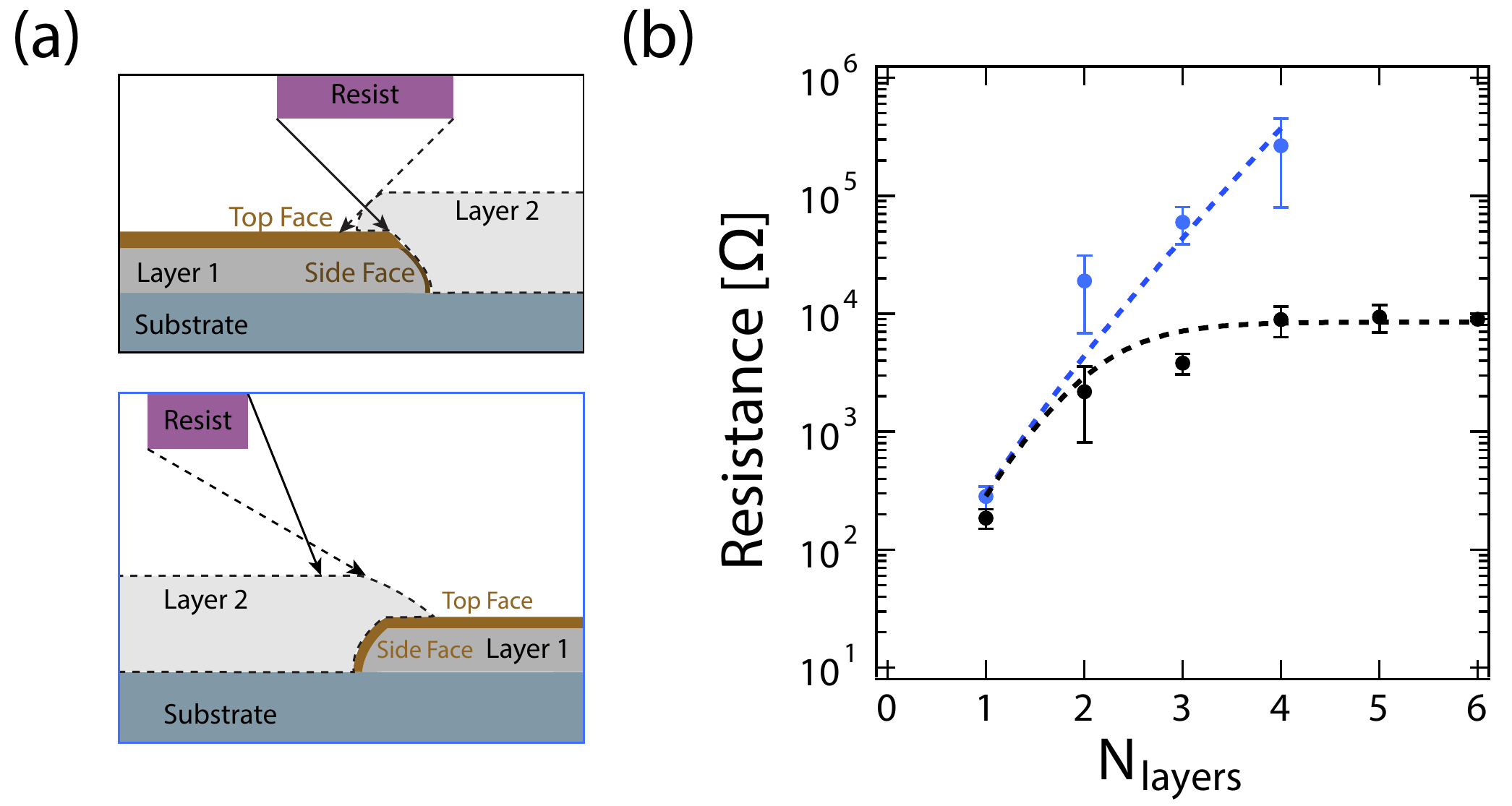}
		\caption{(a) Sketch of the junction formed from metalization and oxidation at different incident angles. When electrodes are evaporated at opposing angles (top, black), filler layers do not contribute to the junction. However, when the electrodes are evaporated from the same direction (bottom, blue), filler layers grow the junction. 
		(b) Normal state resistance of devices with varying number of additional oxide layers ($N_{\rm layers}$) under two separate evaporation/oxidation processes. When evaporated at $\pm45^\circ$ (black), two junctions form. One junction grows slowly, dominating the net resistance. When evaporated at $+24^\circ$ and $+60^\circ$ (blue), a single junction of uniform thickness grows, yielding an exponential increase of the tunnel barrier's resistance. Dashed lines indicate fits to either a one-component (blue) or two-component model (black). Error bars indicate standard deviations of $\sim$25 devices made in each batch.}
		\label{resistanceScaling}
	\end{figure}

	Our oxidation scheme creates two {JJ}s between the electrodes.
	The first {JJ} forms at the bottom electrode's side face, whereas the second {JJ} forms at the top face
	(Fig.~\ref{resistanceScaling}(a)).
	The two {JJ}s have significantly different participation because of their disparate sizes.
	Evaporation at $\pm45^\circ$ results in filler layers only contributing to the top-face JJ.
	The top-face {JJ} has a large area and is relatively thick, due to the added filler layers.
	However, the side-face {JJ} has an area comparable to eBL {JJ}s: its dimensions are 30 nm (the thickness of the bottom electrode) by 1.5~\um~(the length of the Dolan bridge).
	The side-face {JJ} is relatively thin, due to the self-limited growth of aluminum oxide \cite{Cabrera1949,Atkinson1985, Gorobez2021}.
	Because the normal resistance (and the inductance) of a tunnel barrier scales exponentially with the barrier's thickness \cite{Simmons1963,Wang2013, Kang2014, Zeng2015}, the participation of the thick, large-area JJ is $\sim 1\%$ \cite{Minev2020}. The thin, small-area {JJ} dominates the circuit's inductance.
	
	Room-temperature normal state resistance measurements verify the presence of two {JJ}s.
	Figure \ref{resistanceScaling}(b) displays the normal-state-resistance versus the number of additional oxide layers.
	We observe an initial exponential increase that asymptotes to a steady-state value.
	The steady-state value results from the self-limited thickness of aluminum oxide \cite{Cabrera1949, Atkinson1985, Gorobez2021}.
	The measurements are consistent with a parallel-component resistance model (Fig.~\ref{resistanceScaling}(b)).
	
To confirm this model, we also characterize \kater{the normal state resistance of JJs formed} with evaporations on the same side of the Dolan bridge (i.e.~with $+24^\circ$ and $+60^\circ$ instead of $\pm45^\circ$).
	In this configuration, filler layers cover both top- and side-faces of the bottom electrode, allowing equal growth of both {JJ}s.
	For this case, Figure \ref{resistanceScaling}(b) shows an exponential increase in normal-state resistance with the number of oxide layers, consistent with a single {JJ} of constantly increasing thickness \cite{Simmons1963}.

		\begin{figure}
		\centering
		\includegraphics[width=\linewidth]{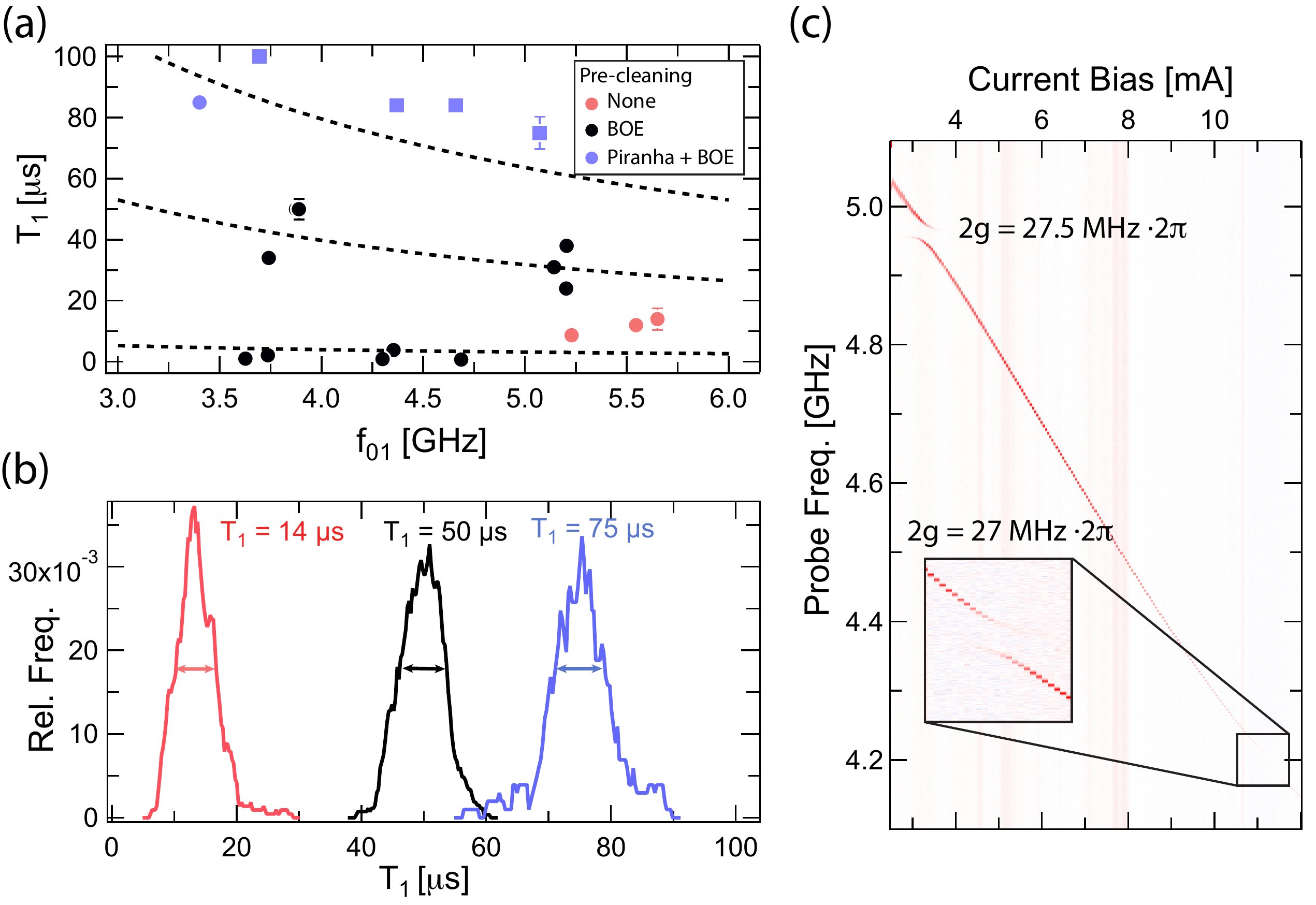}
		\caption{(a) $T_1$ values for different substrate cleaning treatments. 
		    Devices represented as red points received no surface cleaning prior to spinning resist. Devices represented as black points were cleaned with a BOE before spinning and before evaporation. Devices represented as blue points were first cleaned with Piranha, then BOE as above.
			Circular markers received multiple oxidation layers, and square markers received a single, long oxidation.
			Dashed lines indicate constant quality factors, $Q= \omega_{01} T_1$, of 2, 1, and 0.1 million (top to bottom).
			Error bars on select points indicate standard deviations for hours-long repeated measurements, as in panel b.
			(b) Histograms of repeated $T_1$ measurements for each of three cleaning procedures.  Measurements are repeated every 30 seconds over the course of two hours. Averages  of $T_1$ measurements are indicated for each cleaning method.
			(c) An example of avoided crossings as a qubit transition frequency comes into resonance with a strongly coupled TLS. The coupling strength is given by the maximal separation, $2g$, as indicated. \jonathan{The qubit is the 14 $\mu$s sample from panel b (maximum frequency 5.650 GHz, see SI).}
		}
		\label{figT1}
	\end{figure}	
	
	\kater{Room temperature resistance measurements also allow us to study the wafer scale uniformity of the fabrication process.  Photolithography has demonstrated feature-size repeatability down to less than 2\% relative standard deviation \cite{Kim2018,Miller2019,Foroozani2019}. 
	For qubit fabrication, careful treatments of every processing step, including lithography, development, ashing, and oxidation has shown relative standard deviations in resistance of 3.5\% \cite{Kreikebaum2019}.
	For our process, we have optimized to 8\% variation in resistance across a 2-inch wafer.
	We identified non-uniformity of the resist stack as a major source of this variation.
	In particular, by fabricating 16 junctions within a $\sim100\times100 ~{\rm \mu m^2}$ area and repeating this pattern across the wafer, we separate wafer-scale variation from local repeatability, which exhibited a variation of 5\%.}
	 
\kater{Having demonstrated that these fabrication methods produce normal state resistances indicative of JJs appropriate for transmon qubits we now turn to measurements of device coherence times, where we study devices JJs fabricated with $\pm45^\circ$ evaporations and either single layer or multi layer oxidation steps.}

	In Figure \ref{figT1}(a), we show $T_1$ measurements for multiple devices fabricated with either (i) no substrate cleaning (besides plasma ashing), (ii) BOE cleaning before spin coating and before evaporation, or (iii) Piranha and BOE cleaning before spin coating and BOE cleaning before evaporation. The Piranha solution's strong reaction to organic materials precludes cleaning of the developed resist stack.
	We see nearly a five-fold increase in coherence times of devices with multiple  cleaning steps compared to devices without cleaning. 
	 For devices fabricated under various cleaning procedures, we monitor $T_1$ over hours-long timescales \cite{Barends2013, Klimov2018a}. Figure~\ref{figT1}(b) shows time-stability histograms of $T_1$ for these devices. Despite increases in  $T_1$ times with cleaning, we observe significant time variability of $T_1$, which is a measure of the density of weakly coupled TLS fluctuators \kater{\cite{Klimov2018a, schl19, Burnett2019} indicating that there are opportunities to further reduce the TLS density and improve $T_1$}.
	 
	 We utilize two-tone spectroscopy to probe the transmon frequency as a function of applied flux. The presence of strongly coupled TLS defects results in an avoided crossing in the transmon spectrum (Fig.~\ref{figT1}c). Using such measurements, we can directly count the number of strongly coupled TLS defects.  For devices with no substrate cleaning, we observe an average of 1 TLS per device, with each device measured over a range $\sim 1$ GHz.  
	 For 3 out of 4 devices with substrate cleaning, we observed no strongly coupled TLS over a frequency range of $\sim 1$ GHz/device.
	 However, the fourth device ($T_1=84\ \mu$s) with substrate cleaning showed 6 defects in a 600 MHz range. 
	 While substrate cleaning exhibits a clear improvement in average $T_1$ times, we did not observe a significant change in the average number of strongly-coupled TLS fluctuators.

	 %

	\kater{Quarter wave coplanar waveguides (CPWs) were used to minimize sources of loss originating from the materials that were used in transmon qubits in this study. Prior to beginning production of the transmon qubits, microwave CPWs were fabricated to test materials and fabrication techniques. Native oxide was removed from the Si (100) substrate by a 3-minute BOE prior to evaporation of Al \cite{Wisbey2010}. Aluminum films were patterned by selective etching with aluminum etchant type A (Transene Co.) at 50 C.  A vector network analyzer was used to measure the transmission, $S_{21}$, of CPWs using the same experimental setup as discussed previously \cite{zmui12,Wisbey2010}. Resonators were designed \cite{Wisbey2014} with center width of 3 $\mu$m and gap of 2 $\mu$m, to maximize TLS interaction with the device \cite{Gao2008}. The internal quality ($Q_i$) factor, as a function of power, was fit to extract filling-factor-adjusted loss tangent $F \tan\delta_\mathrm{TLS}$,  where $F$ is the filling factor and $\tan\delta_\mathrm{TLS}$ is the internal loss tangent of the material where the TLSs are located. Pappas et al. \cite{Pappas2011} discuss the method for extracting $F \tan\delta_\mathrm{TLS}$ using a power fit of $Q_i$ in detail. High power measurements of the evaporated Al resonators yielded a $Q_i$ of over $1.2\times10^6$ and a lower power $Q_i$ of over $3.0 \times10^5$. 
    We extract $F \tan\delta_\mathrm{TLS} = 2.6\times10^{-7}$ from a fit of $1/Q_i$ to the power-sweep data\cite{Pappas2011,OConnell2008,Wisbey2010}.
	}

		
	Quasiparticle  tunneling is another dominant loss mechanism in state-of-the-art transmon qubits \cite{Martinis2009,Catelani2012,Wang2014,Vool2014,Serniak2019a}.
	Despite our use of infrared-photon filters \cite{Corcoles2011,Barends2011}, our qubits tend to have high effective temperatures ($\sim80\ {\rm mK}$), indicating imperfect high-frequency shielding \kater{(See SI for details on qubit thermometry)}.
	However, we observe that removing the shielding and other lossy filters does not significantly affect $T_1$ times, indicating that \kater{quasiparticles} do not yet significantly limit these devices.
	
	\kater{Our study has focused on characterizing and optimizing the $T_1$ time of fabricated qubits, however the coherence time $T_2$ is also a critical factor in qubit performance. We have evaluated the Ramsey decay $T_2$ time for select devices attaining high $T_1$ times, achieving $T_2^* = 19\ \mu$s and $T_2^\mathrm{e} = 25 \ \mu$s under optimized line filtering (See SI for details on measurement filtering and device parameters). Further improvement of $T_2$ will likely require better filtering to remove intracavity photons \cite{Sears2012,Wang2019}, and investigate the role of vortex trapping  \cite{Stan2004,Song2009,Nsanzineza2014,Wang2014}.}


We have demonstrated a method for fabricating JJs with large lithographic areas. \kater{We have studied coherence of devices based on varying oxidation schemes that achieve suitable critical current densities. One scheme utilizes multiple oxidation deposition steps to create a multi-junction device with small junction areas similar to electron-beam lithography processes, and an alternate scheme utilizes a single long oxidation step. Highest coherence devices utilized a single long oxidation step.}   With appropriate surface cleaning, devices exhibit long \kater{energy decay} coherence times.
	Our work therefore demonstrates photolithography as a viable fabrication approach for high-coherence superconducting quantum devices.
	
	\section*{Supplemental material}
	
Supplemental material details qubit parameters and the measurement setup as well as further information on qubit thermometry.	
	
\begin{acknowledgments}
This research was supported by NSF Grant No. PHY-1752844 (CAREER), the John Templeton Foundation Grant No. 61835, and used of facilities at the Institute of Materials Science and Engineering at Washington University.  
\end{acknowledgments}

The data that support the findings of this study are available from the corresponding author upon reasonable request.

%

		\pagebreak
		
		\textcolor{white}{.}
		
		\pagebreak
		
	\begin{center}
		\textbf{\large Supplemental Material for: ``Optical Direct Write of Dolan\kater{--Niemeyer}-Bridge Junctions for Transmon Qubits''}
	\end{center}

	\makeatletter
	\renewcommand{\theequation}{S\arabic{equation}}
	\renewcommand{\thefigure}{S\arabic{figure}}
	\renewcommand{\bibnumfmt}[1]{[#1]}
	\renewcommand{\citenumfont}[1]{#1}

\subsection{Qubit thermometry}

\kater{We measure the effective temperature of the qubits using the Rabi driving technique introduced in \cite{geer13}. The analysis comprises two measurements. First, we apply resonant driving on the $\{\ket{e},\ket{f}\}$ transition. At low temperature, we assume that the population of $\ket{f}$ is negligible. The resulting oscillation contrast of the excited state probability is proportional the excited state population at the start of the driving. Second, starting from thermal equilibrium, we apply a resonant $\pi$ pulse to transfer population from the ground $\ket{g}$ to excited state $\ket{e}$ of the qubit followed by resonant driving on the $\{\ket{e},\ket{f}\}$ transition. Comparing the Rabi oscillation contrasts of the first and second measurements yields the equilibrium ratio of excited and ground state populations, allowing us to infer an effective temperature based on Boltzmann factors.}

\subsection{Device parameters}
See Supplemental Table \ref{table} for treatments and measurements of each qubit presented in this study.

\begin{table*}
\renewcommand\tablename{Table}
\renewcommand\thetable{S1} 
\kater{\caption{Treatments and measurements of each qubit included in this study. The input setup lists the attenuation at the 4K, 1K, and MC stages of the dilution refrigerator as well as the existence of Eccosorb dissipative filters (E) or K\&L (K) low pass filters. All output lines consist of three circulators and a K\&L low pass filter (MC stage) before a HEMT amplifier at 4K. $\kappa_c$ refers to the cavity linewidth (full width half maximum), $\chi_c$ is the dispersive loading of the cavity, determined by the frequency shift of the cavity resonance from low to high power. }
\begin{tabular}{|p{3.7cm}||p{1.4cm}|p{1.2cm}|p{1.8cm}|p{1cm}|p{1cm}|p{1cm}|p{1cm}|p{1cm}|p{1cm}|p{.8cm}|p{.8cm}|}
Device name              & Oxidation & Cleaning &Input Setup& $f_{c}$ [GHz]&  $\kappa_c/2\pi$ [MHz] &$\chi_c/2\pi$ [MHz]&$f_{01}$ [GHz] & $E_c$ [MHz] & $E_J$ [GHz] & $T_1$ [\us] & $T_2^{\rm E}$ [\us] \\
\hline
JTM.010918.C3D9             & Multiple    & {None} &10/20/30,E,K  &5.874&2.1&  1.0 & 5.228     & 152      & 23.80         & 8.7      & \none        \\
JTM.010918.C5D13            & Multiple    & {None}  &10/20/30,E,K &6.45&1.7 &3.0& 5.545     & 292      & 14.59        & 12       & 3         \\
JTM.012519.C7D10            & Multiple    & {None}  &10/20/30,E,K &5.857&0.4 &2.2& 5.650     & 488      & 9.65         & 14       & 1.3       \\
JTM.041519.E12              & Multiple    & {BOE}  &10/20/30,E,K &5.823&0.7 &1.0& 3.689     & 244      & 7.92        & \none       & \none        \\
DVKJTM.200205.C2D16 v1      & Multiple    & {BOE}   &20/10/30,E,K &5.636&3.8&1.5& 3.874     & 249      & 8.53        & 50       & 35        \\
DVKJTM.200205.C5D23 v1      & Multiple    & {BOE}  &20/20/20,E &5.826&-&-& 3.739     & 266      & 7.54        & 2.6      & \none        \\
DVKJTM.200917.C2D15 v1      & Multiple    & {BOE}  &20/20/20,E &6.869&1.2&2.5& 5.204     & 234      & 15.80         & 38       & \none        \\
DVKJTM.200917.C2D22 v1      & Multiple    & {BOE}  &20/20/20,E&5.819&2.0 &2.7& 4.354     & 234      & 11.24        & 3.8      & 2.2       \\
KZDVK.023062.1B             & Multiple    & {BOE}  &20/20/20,E &6.742&0.6&2& 3.984     & 172      & 12.55        & \none       & \none        \\
KZDVK.023062.1D             & Multiple    & {BOE}   &20/10/10,E &-&- &-& 3.626     & 186      & 9.76        & 1        & 0.5       \\
DVKJTM.200917.C2D03         & Multiple    & {BOE}    &20/20/20,E &5.836&0.3&-& 4.686     & 226      & 13.34        & 0.7      & \none        \\
DVKJTM.200205.C2D16 v2      & Multiple    & {BOE}    &20/20/20,E &5.636&2.6&1.0& 3.889     & 249      & 8.59        & 50       & 10        \\
DVKJTM.200205.C5D23 v4      & Multiple    & {BOE}   &20/20/20,E &5.826&1.0 &-& 3.741     & 256      & 7.80        & 34       & 7.9       \\
DVKJTM.200205.C5D23 v5      & Multiple    & {BOE}    &20/20/20,E&5.852&2.5 &-& 3.736     & 256      & 7.78        & 2.1      & 1.9       \\
DVKJTM.200917.C2D15 v2      & Multiple    & {BOE}   &20/20/20,E &6.869&1.2&2.5 & 5.202     & 234      & 15.79        & 24       & 2.9       \\
DVKJTM.200917.C2D15 v3      & Multiple    & {BOE}   &20/20/20,E &6.869&0.4&1.8 & 5.142     & 234      & 15.44        & 31       & 4.6       \\
DVKJTM.200917.C2D22 v2      & Multiple    & {BOE}    &20/10/10,E &-&-&2& 4.300     & 234      & 10.98        & 0.9      & 0.15      \\
KZDVK\_0400075\_R1\_7       & Single      & {BOE+Pir.} &20/10/10,E &6.869&1.0 &4.5& 5.071     & 280      & 12.78        & 75       & 4.4       \\
KZDVK\_0400075\_C2\_4       & Single      & {BOE+Pir.} &20/10/10,- &5.827&4.0&2.3& 3.695     & 320      & 6.30        & 100      & 1.2       \\
KZDVK\_043080\_R3\_3        & Multiple    & {BOE+Pir.} &20/10/10,E,E &5.827&-&-& 3.401     &\none     &  \none             & 85       & \none        \\
KZ\_047\_R1\_10             & Single    & {BOE+Pir.}  &20/10/10,E,E &5.86&0.6 &5.9 & 4.660     & 294      & 10.43        & 84       & 4         \\
KZ\_047\_R1\_11 v1          & Single    & {BOE+Pir.}  &20/20/20,E &5.839&0.3 &4.5& 4.370     & 296      & 9.19        & 104      & 7.2       \\
KZ\_047\_R1\_11 v2          & Single    & {BOE+Pir.}    &20/20/20,E&5.839&-&- & 4.370     & 296      & 9.19        & 84       & 25.4   
\end{tabular}
\label{table}}
\end{table*}


\end{document}